\documentclass[12pt]{article}
\usepackage{graphicx}

\input{epsf}
\usepackage{cite}
\def\@fmsl@sh#1#2#3{\m@th\ooalign{$\hfil#1\mkern#2/\hfil$\crcr$#1#3$}}
 \def\eq#1\en{\begin{equation}#1\end{equation}}
\def\s[#1,#2]{[#1\stackrel{\star}{,}#2]}
\def\sx[#1,#2]{[#1\stackrel{\star_{x}}{,}#2]}

\newcommand{\nc}{\newcommand}
\nc{\beq}{\begin{equation}}
\nc{\eeq}{\end{equation}}
\nc{\beqa}{\begin{eqnarray}}
\nc{\eeqa}{\end{eqnarray}}

\def\gsim{\mathrel{\rlap{\lower4pt\hbox{\hskip1pt$\sim$}}
    \raise1pt\hbox{$>$}}}       

\begin{document}
\makeatletter
\def\fmslash{\@ifnextchar[{\fmsl@sh}{\fmsl@sh[0mu]}}
\def\fmsl@sh[#1]#2{%
  \mathchoice
    {\@fmsl@sh\displaystyle{#1}{#2}}%
    {\@fmsl@sh\textstyle{#1}{#2}}%
    {\@fmsl@sh\scriptstyle{#1}{#2}}%
    {\@fmsl@sh\scriptscriptstyle{#1}{#2}}}
\def\@fmsl@sh#1#2#3{\m@th\ooalign{$\hfil#1\mkern#2/\hfil$\crcr$#1#3$}}
\makeatother


\title{\large{\bf A Time Variation of  Proton-Electron Mass Ratio and Grand Unification}}

\author{Xavier~Calmet\thanks{xcalmet@ulb.ac.be} \\
Service de Physique Th\'eorique, CP225 \\
Boulevard du Triomphe \\
B-1050 Brussels \\
Belgium \\
and\\
Harald Fritzsch\thanks{fritzsch@mppmu.mpg.de } \\
Sektion Physik, Universit\"at M\"unchen \\                                      
Theresienstr. 37A \\             
D-80333 M\"unchen \\
Germany    
}

\date{May, 2006}

\maketitle

\begin{abstract}

Astrophysical observations indicate a time variation of the proton-electron mass ratio and of the fine-structure constant. We discuss this phenomenon in models of Grand
Unification. In these models a time variation of the fine-structure constant and  of
the proton mass are expected, if either the unified coupling constant
or the scale of unification changes, or both change. We discuss in
particular the change of the proton mass. Experiments in Quantum Optics
could be done to check these ideas.
\end{abstract}

 \newpage

 
A new indication that fundamental constants of nature could evolve with the cosmological time has recently been published \cite{petitjeanPRL}. The ratio of the proton mass to electron mass $\mu=m_p/m_e$ was bigger in the past and has decreased over the last 12 Gyr. The authors of \cite{petitjeanPRL} report
\begin{eqnarray} \label{exinputnew}
\frac{\Delta \mu}{\mu}= (2.4 \pm 0.6) \times 10^{-5}
\end{eqnarray}
which after the result of Webb {\it et al.} 
\begin{equation} \label{exinput}
\frac{\Delta \alpha}{\alpha} = (-0.72 \pm 0.18) \times 10^{-5}
\end{equation}
for a redshift $z \approx 0.5 \ldots 3.5$ \cite{Webb:2000mn},
which indicates a possible
time dependence of the fine structure constant $\alpha$, is the second indication of a possible time dependence of fundamental parameters of nature. We note however that the result of Webb {\it et al.}  could not be reproduced by similar observations \cite{Srianand:2004mq,Levshakov:2005ik,Levshakov:2005ab,Chand:2006va,Quast:2003qu}.

In \cite{Calmet:2001nu,Calmet:2002ja,Calmet:2002jz} (see also \cite{Dent:2001ga,Landau:2000cc,Wetterich:2003jt,Olive:2002tz,Campbell:1994bf,Flambaum:2006ip}  for related works) we used (\ref{exinput}) to predict the parameter $\frac{\Delta \mu}{\mu}$ within the framework of a simple unified theory and found in the simplest case that a time variation of the unification scale could give rise to an effect of the order of $10^{-4}$. In this letter we want to point out that a variation of the unification scale combined with that of the unification coupling constant and/or of the supersymmetry breaking scale could easily lead to an effect of the order of $10^{-5}$ as observed in \cite{petitjeanPRL} . 

As in \cite{Calmet:2001nu,Calmet:2002ja,Calmet:2002jz}, we consider as an example the supersymmetric extension of the standard model. This model can be unified in the framework of SO(10). Supersymmetry is not required to have the unification of the gauge couplings (see e.g.  \cite{Lavoura:1993su,Calmet:2004ck}), but it offers a simple model for our discussion.
We work in the one loop approximation. Time variations of Yukawa and Higgs boson masses can be neglected within this approximation. Furthermore not all time variations of physical scales can be observables, only ratios of scales or dimensionless quantities can be observed.  Within these approximations, we only have three parameters: the unification scale $\Lambda_u$,  the unified coupling constant $\alpha_u$ and the scale of the supersymmetry breaking $\Lambda_{S}$.  We note that the proton mass is determined mainly by the QCD scale. Quark masses do not play a big role. Furthermore, QED splits the neutron and proton masses, besides the quark masses, and this ratio could be time dependent and lead to an observable effect .

We now focus on the QCD scale $\Lambda_{QCD}$ and extract its value from the Landau pole of the renormalization group equations for the couplings of the supersymmetric standard model
\begin{eqnarray} \label{running}
  \alpha_3(\mu)^{-1}
 &=&
  \left ( \frac{1}{\alpha_u(\Lambda_{u})}+\frac{1}{2 \pi}
  b^{S}_3  \ln
  \left ( \frac{\Lambda_{u}}{\mu} \right) \right) \theta(\mu- \Lambda_{S})
\\ \nonumber &&  +
\left ( \frac{1}{\alpha_3(\Lambda_{S})}+\frac{1}{2 \pi}
  b_3   \ln
  \left ( \frac{\Lambda_{S}}{\mu} \right) \right)
\theta(\Lambda_{S}-\mu),
\end{eqnarray}
where the parameters $b_i$ are given by $b_i\!\!\!=(b_1,
b_2, b_3)=(41/10, -19/6, -7)$  and by $b^{{S}}_i\!\!\!=(b^{{S}}_1, b^{{S}}_2, b^{{S}}_3)=(33/5,
1, -3)$ and where
${\alpha_i(\Lambda_{S})}$ is the value of the coupling constant at
the supersymmetry breaking scale which can be expressed in terms of $\Lambda_S$, $\alpha_u$ and $\Lambda_u$ \footnote{We would like to point our that our formalism is not dependent on whether nature is supersymmetric or not at some higher scale. Indeed the same formalism would apply to an SO(10) gauge symmetry broken to the standard model with an intermediate gauge symmetry. In that case the $b^S_i$ would be the beta-functions of the intermediate gauge symmetry and $\Lambda_S$ the energy scale of the intermediate gauge symmetry breaking.}. The QCD scale is given by
\begin{eqnarray}
\Lambda_{QCD}= \Lambda_S \left(\frac{\Lambda_u}{\Lambda_S}\right)^{\frac{b_3^S}{b_3}} \exp\left(\frac{2 \pi}{\alpha_u}\right)^{\frac{1}{b_3}}.
\end{eqnarray}
The time variation of $\Lambda_{QCD}$ is then determined by
\begin{eqnarray}
\frac{\dot \Lambda_{QCD}}{\Lambda_{QCD}}= -\frac{2\pi}{b_3} \frac{\dot \alpha_u}{\alpha_u^2}
+\frac{b_3^S}{b_3} \frac{\dot \Lambda_u}{\Lambda_u} + \frac{b_3-b_3^{S}}{b_3} \frac{\dot \Lambda_S}{\Lambda_S}.
\end{eqnarray}
This equation determines the ratio $\frac{\Delta \mu}{\mu}$, since we keep the electron mass constant. We thus find
\begin{eqnarray} \label{result}
\frac{\Delta \mu}{\mu}= -\frac{2\pi}{b_3} \frac{\dot \alpha_u}{\alpha_u^2}
+\frac{b_3^S}{b_3} \frac{\dot \Lambda_u}{\Lambda_u} + \frac{b_3-b_3^{S}}{b_3} \frac{\dot \Lambda_S}{\Lambda_S}= \frac{2 \pi}{7}  \frac{\dot \alpha_u}{\alpha_u^2} +\frac{3}{7}  \frac{\dot \Lambda_u}{\Lambda_u}-\frac{4}{7} \frac{\dot \Lambda_S}{\Lambda_S}.
\end{eqnarray}
This equation is rather instructive. A cancellation between the functions $ \frac{\dot \alpha_u}{\alpha_u^2}$, $ \frac{\dot \Lambda_u}{\Lambda_u}$ and  $\frac{\dot \Lambda_S}{\Lambda_S}$ could easily occur and give  $\frac{\Delta \mu}{\mu}\sim 2 \times 10^{-5}$. If we assume the result eq. (\ref{exinput}) for the fine-structure
constant and take only a time dependence of the unification scale or of
the unified coupling constant in account, we find \cite{Calmet:2002ja} for the  mass ratio
either $\Delta \mu/ \mu \sim 22\times10^{-5}$ (for $\alpha_u=$ const.) or $\Delta \mu/\mu \sim -27\times10^{-5}$ (for $\Lambda_u=$ const.). Thus one can obtain the observed result  (\ref{exinputnew}) by having a cancellation in eq. (\ref{result}).

It is interesting to point out that the measurement (\ref{exinput}) provides a direct determination of the time dependence of the unified coupling constant:
\begin{eqnarray}
\frac{\dot \alpha_u}{\alpha_u^2}=\frac{3}{8}\frac{\dot \alpha}{\alpha^2}
\end{eqnarray}
since this relation is renormalization scale invariant. We find $\dot\alpha_u/\alpha_u^2=-3\times 10^{-14} \mbox{yr}^{-1}$ using  (\ref{exinput}) as an input, i.e. the unified coupling constant was smaller in the past. If we assume that the supersymmetry scale is time independent, we find that the new measurement (\ref{exinputnew}) together with (\ref{exinput}) allows to extract the time variation of the unification scale:
\begin{eqnarray}
\frac{ \dot \Lambda_u}{\Lambda_u} = \left (\frac{\dot \Lambda_{QCD}}{\Lambda_{QCD} }+ \frac{2 \pi}{b_3} \frac{3}{8} \frac{\dot \alpha}{\alpha^2}  \right) \frac{b_3}{b^S_3}.
\end{eqnarray}
We find $\dot \Lambda_u/\Lambda_u=7\times 10^{-14}\mbox{yr}^{-1}$, i.e. the unification scale was higher in the past.

The new observation (\ref{exinputnew}) implies a time variation for the proton-electron mass ratio $\frac{\Delta{\mu}}{\mu}$ of the order  of $2 \times 10^{-15} \mbox{yr}^{-1}$, if linearly extrapolated, which should be observable in quantum optics experiments using modern techniques. This is in contrast to the expectation of  \cite{Calmet:2002ja} where a time variation of the grand unification scale only would imply a change of $\frac{\Delta{\mu}}{\mu}\sim 3 \times 10^{-14} \mbox{yr}^{-1}$ which is now experimentally excluded.

A time variation of $2 \times 10^{-15}$ per year can be observed by precise
experiments in quantum optics, e.g. by comparing a cesium clock with
hydrogen transitions, as done in ref. \cite{haensch}. In a cesium clock the time is measured by using a hyperfine transition. The frequency of the clock depends therefore on the magnetic moment of the
cesium nucleus. The latter is directly proportional to $\Lambda_{QCD}$. The hydrogen transitions, however, are only dependent on the electron mass, which we  assume to be constant.

A time variation of the QCD scale and of the unification scale of the order discussed above would imply that considerable changes of physics are expected at times very close to the Big Bang.  For example, the results for nucleosynthesis will be  changed (see e.g. \cite{Kolb:1985sj,Nollett:2002da}). However the details are highly model dependent and beyond the scope of this paper.

\bigskip
\subsection*{Acknowledgments}
\noindent 
This work was supported in part by the IISN and the Belgian science
policy office (IAP V/27).

\bigskip


\end{document}